# A 1.2-V 162.9-pJ/cycle Bitmap Index Creation Core with 0.31-pW/bit Standby Power on 65-nm SOTB

Xuan-Thuan Nguyen*, Trong-Thuc Hoang*, Hong-Thu Nguyen*, Katsumi Inoue†, and Cong-Kha Pham*

*Abstract*—The ability to maximize the performance during peak workload hours and minimize the power consumption during off-peak time plays a significant role in the energy-efficient systems. Our previous work has proposed a high-performance multi-core bitmap index creator (BIC) in a field-programmable gate array that could deliver higher indexing throughput than central processing units and graphics processing units. This brief extends the previous study by focusing on the application-specific integrated circuit implementation of the proposed BIC in a 65-nm silicon-on-thin-buried-oxide (SOTB) CMOS process. The BIC chip can operate with different supply voltage from 0.4 V to 1.2 V. In the active mode with the supply voltage of 1.2 V, the BIC chip is fully operational at 41 MHz and consumes 162.9 pJ/cycle. In the standby mode with the supply voltage of 0.4 V and clock-gating technique, the power consumption was reduced to 10.6 $\mu$W. The standby power is also dramatically reduced to 2.64 nW due to the utilization of reverse back-gate biasing technique. This achievement is considerable importance to the energy-efficient systems.

## I. INTRODUCTION

Analyzing a considerable amount of data generated by scientific observations and experiments is undoubtedly a compute-intensive and time-consuming task. To reduce the analytics time, bitmap index (BI) [1] is utilized because it allows the multi-dimensional queries, which account for the majority of the total queries, to be efficiently answered by simply using the bitwise logical operations. BI also strongly supports parallel processing, which makes it suitable for the modern platforms, such as graphics processing units (GPUs) and central processing units (CPUs). However, the performance of GPUs and CPUs mainly depends on the number of cores usage. For example, a BI system introduced by T. Zhong *et al.* [2] could produce an indexing throughput of 108 MB/s and 473 MB/s if the number of CPU cores are 16 and 60, respectively. The more the cores are exploited, the higher the power consumption increases.

Because field-programmable gate arrays (FPGAs) and application-specific integrated circuits (ASICs) can achieve not only higher computing efficiency but also lower power consumption compared to both GPUs and CPUs, they have been increasingly important to resolve the compute-intensive tasks. This was exemplified by an FPGA-based large scale deep neural network system [3], whose processing throughput was around 3.5 and 2.5 times higher and power consumption was four and 22 times lower than the 80-W CPU and 225-W GPU, respectively. Following this research trend, our previous work [4] proposed a 150-MHz FPGA-based BI creation (BIC) system that achieved 2.8 times and 1.7 times higher indexing throughput than CPU- [2] and GPU-based designs [5], respectively. Despite the remarkable achievement of performance, eliminating as much as leakage power during the standby mode is still one of the most challenging tasks.

It is well-known that the standby leakage power significantly increases according to the transistors scaling. This is because in sub-100-nm technologies, it is challenging to further decrease ($V_{DD}$) because of the increase in threshold voltage ($V_{TH}$) variation of CMOS devices. In order to solve this problem, R. Tsuchiya *et al.* [6] proposed a so-called 65-nm silicon-on-thin-buried-oxide (SOTB) technology that can reduce the standby leakage power consumption dramatically by using the reverse back-gate biasing (RBB) technique. As a result, this technology was widely applied in many applications, such as embedded microprocessor [7], embedded memory [8], and content-addressable memory (CAM) [14]. Our previous work also implemented a BIC core in the 65-nm SOTB [16]. The post-layout simulation results proved the ASIC feasibility.

This brief extends our previous work [4], [16] by originally proposing a proof-of-concept BIC chip in a 65-nm SOTB CMOS process. This chip could operate at a wide-range supply voltage between 0.4 V and 1.2 V. At 1.2 V, it was fully operational at 41 MHz and required 6.68 mW. The standby power reduction is performed by clock-gating (CG) and reverse back-gate bias (RBB) techniques. By using CG only, the standby power dropped to 10.6 $\mu$W. However, with the help of RBB, the standby power sharply reduced to 2.64 nW, or 4,027 times. As compared to power-gating (PG) technique, we require no data retention function that need to restore the data of sequential circuits during the standby mode.

The remainder of this paper is organized as follows. Section II briefly presents the concept of bitmap index and SOTB architecture, as well as their related works. Section III briefly summarizes the hardware architecture of BIC system. Section IV shows the chip measurement results in 65-nm SOTB CMOS

Manuscript received ******** ** ****; revised ******** ** ****; accepted ******** ** ****. Date of publication ******** ** ****; date of current version ******** ** ****.

*The authors are with the Department of Engineering Science, the University of Electro-Communications, Tokyo, Japan, e-mail: xuan-thuan@vlsilab.ee.uec.ac.jp. †The author is with Advanced Original Technologies Co., Ltd (AOT), Tokyo, Japan. This work was supported in part by the VLSI Design and Education Center, in part by the University of Tokyo in collaboration with Synopsys, Inc., and in part by the Cadence Design Systems, Inc.

Color versions of one or more of the figures in this brief are available online at http://ieeexplorer.ieee.org.

Digital Object Identifier

|       | $O_1$ | $O_2$ | $O_3$ | $O_4$ | $O_5$ | $O_6$ | $O_7$ | $O_8$ | $O_9$ |
|-------|-------|-------|-------|-------|-------|-------|-------|-------|-------|
| $A_1$ | 1 | 0 | 0 | 1 | 0 | 0 | 0 | 0 | 1 |
| $A_2$ | 0 | 1 | 0 | 0 | 0 | 1 | 1 | 1 | 0 |
| $A_3$ | 0 | 0 | 0 | 0 | 0 | 0 | 0 | 0 | 0 |
| $A_4$ | 1 | 0 | 0 | 0 | 0 | 0 | 0 | 1 | 0 |
| $A_5$ | 0 | 1 | 0 | 1 | 0 | 0 | 0 | 0 | 0 |

A: Attribute O: Object

Fig. 1: An example of bitmap index.

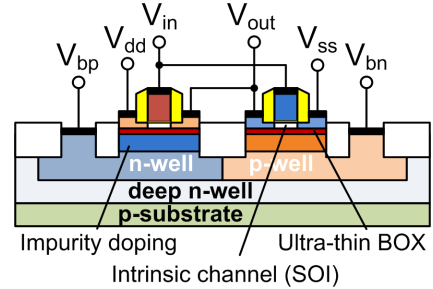

Fig. 2: The structure of SOTB device.

process. Section V finally gives the conclusion.

## II. BACKGROUND

### A. Bitmap Index (BI)

Fig. 1 gives a BI example of nine objects (O) and five attributes (A). The number of columns and rows of BI are identical to the number of objects and attributes. The value BI($i$, $j$) turns into one if $O_j$ contains $A_i$. Otherwise, BI($i$, $j$) is assigned to zero. Each row of BI, therefore, indicates the index of a certain attribute in a set of nine objects. By using this BI, the multi-dimensional queries, such as "*find all objects containing both $A_2$ and $A_4$, but not $A_5$*," can be effectively answered by performing a set of bitwise logical operations ($A_2$ AND $A_4$ AND (NOT $A_5$)). Due to this advantages, BI is widely employed in data warehousing applications, where most queries involve multiple dimensions. This work, however, only focuses on BI creation because it is the most expensive task in query processing [9].

### B. Silicon-on-Thin-Buried-Oxide (SOTB)

Fig. 2 illustrates a structure of a SOTB CMOS device, which is an enhancement of the fully-depleted silicon on insulator (FD-SOI) CMOS. The ultra-thin BOX layer makes the transistor highly immune from the short channel effect, and its intrinsic channel without halo implant suppresses the $V_{th}$ variation caused by the random dopant fluctuation. The BOX layer and impurity doping in the substrate directly under the BOX enable a multiple $V_{th}$ design. They also enable the wide-range back-gate controllability, which allows designers and users to optimize the chip power after it is fabricated. The back-gate bias voltage $V_{bb}$ is defined in Eq. (1).

$$V_{bb} = V_{bn} = V_{dd} - V_{bp} \qquad (1)$$

One of the advantages of SOTB is that it can significantly reduce the standby leakage current ($I_{stb}$) of an integrated circuit. It is reminded that in the bulk CMOS device, $I_{stb}$ mainly consists of the subthreshold leakage current ($I_{slc}$) and gate-induced drain leakage current ($I_{gidl}$) [10]. Due to the SOTB structure, $I_{gidl}$ is suppressed by properly adjusting the length of the overlapped region between gate and source/drain extensions. Additionally, $I_{slc}$ is reduced by effectively applying the reverse $V_{bb}$. In fact, the more the reverse $V_{bb}$ is applied, the more the $I_{slc}$ decreases. For those reasons, SOTB is recognized as a prime candidate for ultra-low standby-power applications.

## III. HARDWARE ARCHITECTURE

### A. Overview

Fig. 3 depicts the architecture of a BIC core, which is composed of a content-addressable memory (CAM), a buffer, and a transpose matrix (TM). This core is used to index $N$ records by $M$ given keys. First, record $R_1$ is fed into CAM. Second, all $M$ keys enter CAM in turn. If $R_1$ contains the incoming key, CAM returns a single 1-bit. Otherwise, it returns 0-bit. Third, each generated bit is sequentially written into the first row of the buffer. As soon as the last key $K_M$ is used, $R_2$ is fed to BIC instantly for the next process. The three-step procedure repeats until the final record $R_M$ is properly indexed. Finally, a TM module swaps all of the rows of the buffer to the correspondent columns, and vice versa. The final result is a $M \times N$-bit BI.

### B. Content-Addressable Memory (CAM)

CAM is a special type of memory used in many high-speed search applications, such as image processing and information retrieval. Although modern FPGAs provide a large number of embedded RAM blocks, dedicated registers, and lookup tables, they exclude dedicated CAM blocks, presumably because of their disadvantages of area and power. Instead, FPGA vendors propose the methodology to construct a scalable binary CAM from the available embedded RAM blocks of FPGA by performing a special mapping technique to input data and address [11]. The proposed CAM is composed of several 32-word×8-bit CAM blocks (CBs), with each one is built by an 8-Kbit dual-port RAM. After receiving a record, each key is sent to CAM in turn. The matching bit is immediately returned in the next clock, i.e. one if the record contains this key and zero in the opposite case.

### C. Buffer

Buffer is used to temporary stored the data from the CAM. It is also made by a set of dual-port RAMs that allow data to be read and written simultaneously. As soon as buffer is full, each row is dispatched to the next module.

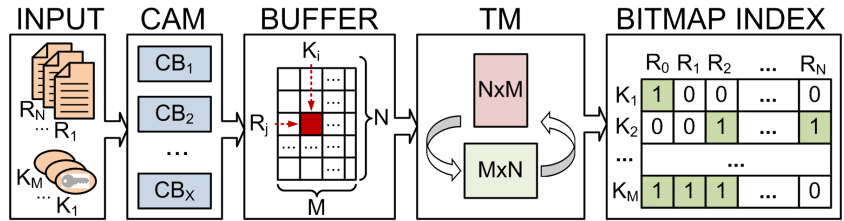

Fig. 3: BIC core.

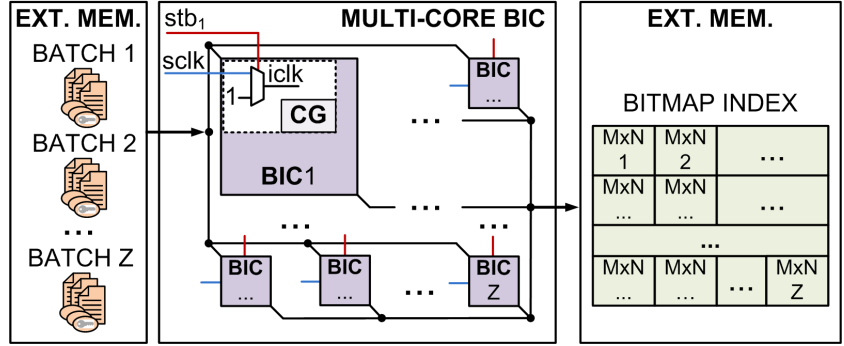

Fig. 4: Multi-core BIC system.

*D. Transpose Matrix (TM)*

TM converts all buffer rows into BI columns. It is composed of a control unit and a transpose unit. The former manages to read inputs from BI and write outputs to BIT while the latter focuses on the matrix transposition. The output of TM is each row of BI.

*E. Multi-core BIC and Power Management*

Fig. 4 introduces the concept of a multi-core BIC system, where $Z$ BIC cores are deployed to index data in parallel. Each set of records and keys is stored as a batch in an external memory in advance. During the operation, batch $i$ is sent to BIC $i$ for indexing. Upon completion, each BI result $i$ are orderly dispatched to the external memory for storage. Depending on the workload, a specific number of BIC cores are activated. The remainders are put into standby mode to save the energy.

To control the dynamic power, a simple clock-gating (CG) circuit is integrated into each BIC core. In the active mode, BIC is directly supplied by the system clock *sclk*. In the standby mode, $stb_1$ turns into one to isolate *sclk* from BIC. Depending on the workload, a certain number of BIC can be turned off to save the dynamic power. The static power, on the other hand, will be handled by RBB technique.

## IV. EXPERIMENTAL RESULTS

The proof-of-concept BIC core including an CAM, a buffer, and a TM, were manufactured in a 65-nm SOTB CMOS technology. Due to the restriction on the chip packet's area, the BIC size had to be adjusted. The number of keys and records were reduced from 16 to 8 keys and 256 to 16 records, respectively. Each record contains 32 8-bit words, instead of 256 words. Accordingly, a 32×8-bit CAM was designed to fix on each record. Because the CAM was built from the RAM, where one CAM cell cost 32 RAM bits [11], 32×32×8 = 8,192 memory bits were required. Furthermore, the buffer module contained a $N{\times}M$-bit matrix, so the memory bits were 16×8 bits = 128 bits. The total number of used memory bits of a BIC core, therefore, was 8,320 bits. It is noted that each memory bit was made by the dedicated register defined in Verilog HDL code.

Fig. 5 illustrates the die photo of proof-of-concept BIC core and its features. The core size is 648×320 $\mu m^2$ and cost around 466,854 transistors. The 65-nm chip could operate at a wide-range supply voltage between 0.4 V and 1.2 V. At 1.2 V, it was fully operational at 41 MHz and consumed 6.68 mW. More significantly, its standby power could reduce up to 4.24 nW, owing to the application of both CG and RBB techniques. It is noted that the operating frequencies of 150 MHz obtained in the post-layout simulations [16] were those of the BIC cores only. In the fabricated chip, the interconnects between the BIC core and chip packet, as well the chip packet itself, contributed to the overall latency. As a result, the measured frequencies were approximately six times slower than the simulated ones.

Fig. 6 shows the variation of frequency when the supply voltage $V_{dd}$ ranges from 0.4 V to 1.2 V. The minimum frequency was 10.1 MHz at 0.4 V, while the maximum frequency reached 41 MHz at 1.2 V. The minimum and maximum power consumptions were 0.17 mW and 6.68 mW, respectively. Fig.

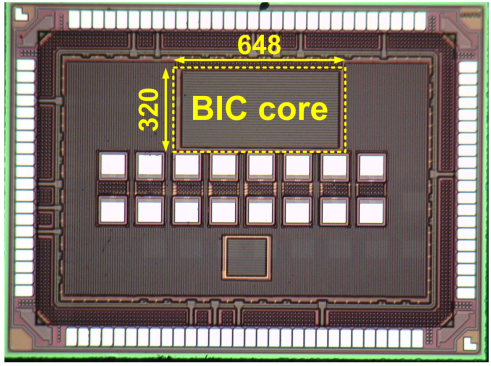

| Technology | 65-nm SOTB CMOS |
|---|---|
| Core size | 0.21 mm² |
| # of Cells | 36,205 |
| # of Transistors | 466,854 |
| I/O $V_{dd}$ | 3.3 V |
| Core $V_{dd}$ | 0.4 V – 1.2 V |
| Active Power (Sim.) | 150 MHz @ 4.69 V @ 0.55 V |
| Active Power | 22 MHz @ 0.6 mW @ 0.55 V<br>41 MHz @ 6.68 mW @ 1.2 V |
| Standby Power | 2.64 nW @ 0.4 V |

*Measured at room temperature*

Fig. 5: Die photo and features.

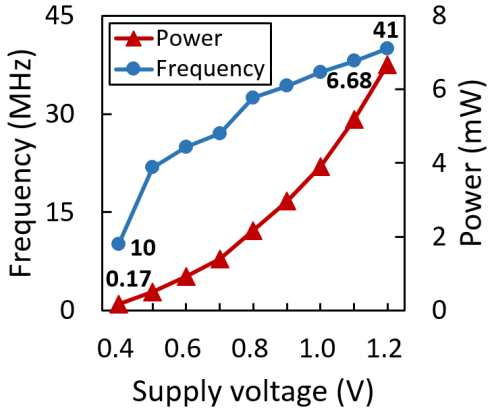

Fig. 6: Die photo and features.

7 depicts the energy consumption of BIC core defined as the quotient of power consumption and frequency. The highest energy point was 162.9 pJ/cycle at 1.2 V.

Fig. 8 shows the leakage current in standby mode $I_{stb}$, when $V_{bb}$ varies from 0 V to -2 V and $V_{dd}$ varies from 0.4 V to 1.2 V. At $V_{dd}$ = 0.4 V, whenever $V_{bb}$ decreases by 0.5 V, $I_{stb}$ is proportionally reduced by one order of magnitude. The minimum $I_{stb}$ = 6.6 nA achieves at $V_{bb}$ = -2 V. It should be noted that at high reverse $V_{bb}$, $I_{stb}$ slightly increases according to the increase in $V_{dd}$. For example, if $V_{dd} > 0.8$ V, $I_{stb}$ of $V_{bb}$ = -2 V exceeded that of $V_{bb}$ = -1.5 V. The reason is that when $V_{dd}$ was small, $I_{gidl}$ was suppressed due to the SOTB structure mentioned earlier. However, if $V_{bb}$ was small and $V_{dd}$ became high, $I_{gidl}$ sharply increased and completely dominated $I_{stb}$ [7].

Table I gives the comparison in the standby power/bit (SPB) between our design and the others which focused on the energy-efficient CAM-based search engines. The key implementation of each design is shown as follows.

- Ref. [12] presented an energy-efficient full-custom CAM design for IPv6 lookup tables in a UMC 65-nm CMOS technology. Two power-gating techniques, namely super cut-off and multi-mode data-retention, were utilized to reduce the leakage power by up to 29.8%.
- Ref. [13] presented an energy-efficient full-custom CAM design in a UMC 40-nm low power platform (LP) CMOS process. The column-based data-aware power control was employed by power-gating devices for up to 59.8% of leakage power reduction.
- Ref. [14] implemented an SRAM-based CAM for fast and exact pattern matching in 65-nm SOTB CMOS process. RBB technique with $V_{bb}$ = -2 V and $V_{dd}$ = 0.4 V was performed to reduce the leakage current by three orders of magnitude.
- Ref. [15] proposed a reconfigurable full-custom CAM/SRAM for parallel data search in 28-nm FD-SOI CMOS technology. According to the authors, the measured leakage current/bit was 4.35 pA at 0.4 V, or equivalent to SPB of 1.74 pW/bit.

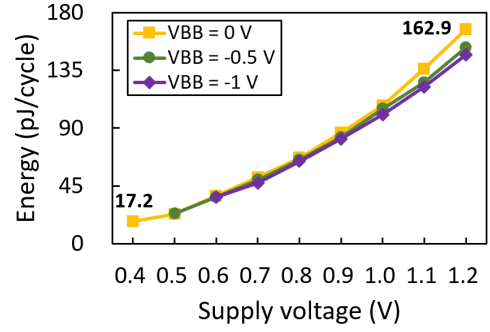

Fig. 7: Energy consumption.

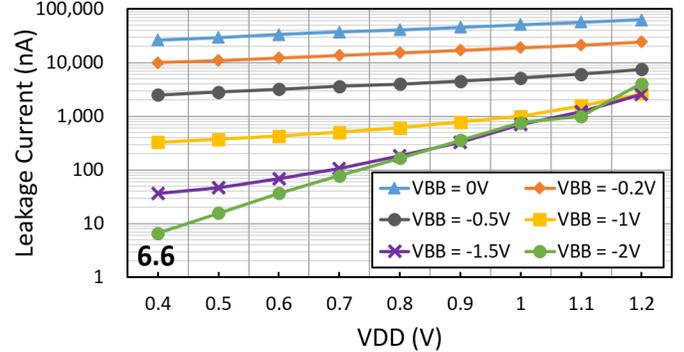

Fig. 8: Energy consumption.

TABLE I: The comparison in SPB.

|  | Ref. [12] | Ref. [13] | Ref. [14] | Ref. [15] | This work |
|---|---|---|---|---|---|
| Technology (nm) | 65 | 40LP | 65SOTB | 28FDSOI | 65SOTB |
| Area ($mm^2$) | 0.43 | 0.07 | 1.60 | 0.33 | 0.21 |
| Memory (Kbits) | 36 | 10 | 64 | 8 | 8.125 |
| Stb. techniques | PG | PG | CG+RBB | – | CG+RBB |
| Stb. power ($\mu$W) | 842 | 201 | 0.12 | – | 0.0026 |
| SPB (pW/bit) | 22,841 | 19,628 | 1.83 | 1.74 | **0.31** |

PG: power gating, CG: clock gating, RBB: reverse back-gate biasing

As seen in Table I, our SPB dramatically dropped to 0.31 pW/bit. As compared to PG techniques, which were presented in Ref. [12] and Ref. [13], our SPB cost only 0.0013% and 0.0016%, respectively. As compared to FD-SOI process, which was proposed in Ref. [15], our SPB cost only 17.8%. Finally, at the same SOTB technology, we outperform Ref. [14] approximately 16.9%.

## V. Conclusion

A proof-of-concept ASIC for data indexing in a 65-nm SOTB CMOS has been originally proposed in this brief. By employing the simple CG and RBB techniques, an ultra-low standby power of 0.31 pW/bit has been attained. This achievement is extremely valuable to the energy-efficient multi-core platforms, where eliminating the standby power of the inactive cores is indispensable.